\documentclass[12pt]{article}
\usepackage{bm}
\usepackage{amssymb,graphicx,color,epsfig}
\newcommand{\be}{\begin{equation}}
\newcommand{\ee}{\end{equation}}
\newcommand{\bear}{\begin{eqnarray}}
\newcommand{\eear}{\end{eqnarray}}
\newcommand{\ba}{\begin{array}}
\newcommand{\ea}{\end{array}}
\newcommand{\nn}{\nonumber}

\def\ket#1{|#1 \rangle}

\begin{document}

\vspace{.9in}
\begin{flushright}
TCDMATH 08--01
\end{flushright}
\vspace{0.5cm}

\vspace{15mm}

\begin{center}
{{{\Large \bf Marginally Deformed Rolling Tachyon around the Tachyon
Vacuum in Open String Field Theory}}\\[15mm]
{O-Kab Kwon}\\[3mm]
{\it School of Mathematics, Trinity College, Dublin, Ireland,}\\[1mm]
{\it Department of Physics and Institute of Basic Science\\
Sungkyunkwan University, Suwon 440-746, Korea }\\[1mm]
{\tt okabkwon@maths.tcd.ie, okab@skku.edu} }
\end{center}
\vspace{15mm}

\begin{abstract}

We investigate the string field theory around the tachyon vacuum. A
pure gauge form of the solution is constructed at the tachyon
vacuum. For a special choice of the gauge function for the pure
gauge form, marginal deformation from the tachyon vacuum is allowed
due to the nontrivial roles of Schnabl's analytic vacuum solution.
We obtain an exact rolling tachyon solution which describes the late
time behaviors of D-brane decay.

\end{abstract}


\newpage

\section{Introduction}

There are two well-known vacua in open bosonic string field theory
(OSFT)~\cite{Witten:1985cc}: the unstable (perturbative) vacuum and
 the tachyon (nonperturbative) vacuum. M. Schnabl obtained the
analytic solution for the tachyon vacuum~\cite{Schnabl:2005gv}.
After Schnabl's work, there has been remarkable progress in
understanding OSFT~\cite{Okawa:2006vm}-\cite{Kiermaier:2007jg}.
Especially, many works have been devoted to the construction of
analytic solutions in bosonic
string~\cite{Schnabl:2007az,Kiermaier:2007ba,
Fuchs:2007yy,Okawa:2007it,Kishimoto:2007bb,Kiermaier:2007vu,Lee:2007ns}
and superstring~\cite{Erler:2007rh,Okawa:2007ri,Kishimoto:2007bb,
Fuchs:2007gw,Kiermaier:2007ki} field theories, which correspond to
the exactly marginal deformations of the boundary conformal field
theory (BCFT)\cite{Callan:1994ub}. For the earlier works on marginal
deformations in OSFT, see
Refs.~\cite{Sen:1990hh,Sen:2000hx,Takahashi:2002ez,
Kluson:2002av,Kluson:2003xu,Sen:2004cq,Kishimoto:2005bs}.

One of the most interesting examples of marginal deformations is the
time dependent solution, referred to as rolling tachyon, in open
string theory. Much of the interest in the rolling tachyon solution,
however, have been concentrated on the deformations from the
unstable vacuum, such as the recently developed marginal
deformations in OSFT. Since the known marginal deformations are
exact but perturbative solutions in the perturbation parameter
$\lambda$ and  the closed forms are not known up to now, the
knowledge of the marginally deformed rolling tachyon in OSFT is
restricted to the physics around the unstable vacuum, and several
puzzles still exist.

The rolling tachyon solutions~\cite{Sen:2002nu,Sen:2002in} in BCFT,
the boundary string field theory (BSFT), and the low energy
effective field theories describe the tachyon matter interpreted as
 closed string radiations from the D-brane decay. During the decay
process of D-brane, which is encoded in the dynamics of the tachyon
field, the pressure of the system approaches zero monotonically from
a negative value, maintaining a constant energy density, and the
tachyon field grows monotonically and approaches the tachyon vacuum,
which is located at infinity of the tachyon field.

In OSFT, however, the different behaviors of rolling tachyon appear
in the level truncated field
theory~\cite{Moeller:2002vx,Fujita:2003ex,Coletti:2005zj}. In the
$(0,0)$-level truncation of $L_0$-eigenstate expansion, the rolling
tachyon solution overshoots the tachyon vacuum and oscillates with
ever-growing amplitude. Moreover, the pressure of the system has
similar oscillating behaviors. The qualitatively similar behaviors
also appear in the $p$-adic string theory
also~\cite{Moeller:2002vx}. These unexpected results (that the
tachyon field does not roll from the unstable vacuum to the tachyon
vacuum) were also confirmed in the higher
level~\cite{Coletti:2005zj,Forini:2006tn}. Even in the exact
marginal deformation for rolling tachyon, these oscillating
behaviors seem to appear~\cite{Schnabl:2007az,Kiermaier:2007ba},
although they are difficult to confirm since the coefficients in the
series expansion for the tachyon solution can be obtained
numerically, except for several ones, and are restricted to few
coefficients.

The aforementioned is a puzzle in the time dependent behaviors of
OSFT. How, then, can the puzzling behaviors of rolling tachyon in
OSFT be reconciled with the well-known behaviors in BCFT, BSFT, and
other low energy effective theories? Several trials were conducted
to find the answer to this question, where the time dependent gauge
transformation~\cite{Coletti:2005zj,Ellwood:2007xr},
other kinds of time dependent solution~\cite{Calcagni:2007wy}
and the properties of the partition function in a two-dimensional
sigma model~\cite{Jokela:2007dq} were used.

In this work, we try to solve the puzzle of the rolling tachyon
solution in OSFT. This is done by considering the rolling tachyon
marginal deformation from the tachyon vacuum. We investigate the
string field theory around the tachyon vacuum background, which is
explicitly known by Schnabl~\cite{Schnabl:2005gv}. The action for
the string field around the tachyon vacuum has the same form as the
action around the unstable vacuum when BRST operator $Q_B$ at the
unstable vacuum is replaced by  BRST operator $\tilde Q$ at the
tachyon vacuum. We construct an analytic solution perturbatively in
a parameter $\lambda$ at the tachyon vacuum, using the remarkable
properties of wedge states with operator
insertions on the worldsheet boundary~\cite{Rastelli:2000iu}.
Since we insert a matter operator on the boundary in the profile of
string field $\tilde\Psi$ obtained in section 3,
it seems that $\tilde\Psi$ is not well-defined
at the tachyon vacuum which has no worldsheet boundary. However, it
does not cause any problem since $\tilde\Psi=0$ represents
the tachyon vacuum by construction.

In section 2, it is shown, by explicitly solving equation of motion
 and appropriate gauge transformations, that the vanishing
cohomology of $\tilde Q$ at the tachyon vacuum is directly connected
to the pure gauge forms for the well-known perturbative type of
solution in OSFT. We argue the validity of the pure gauge solutions
for some cases corresponding to large gauge transformation.

In section 3, we consider the marginal deformations around the
tachyon vacuum. Through a special choice of ghost number zero string
state $\phi$, we construct the marginally deformed solutions. We
apply the marginally deformed solution to the rolling tachyon vertex
operator $e^{-X^0}$, which describes the late time behaviors of
D-brane decay. We obtain the explicit rolling tachyon profile around
the tachyon vacuum. Finally, the conclusions that are arrived at in
this work are presented in section 4.

\section{Pure Gauge Solution around the Tachyon Vacuum}

Let us first briefly review the bosonic OSFT around the tachyon
vacuum. The OSFT around the tachyon vacuum solution is described by
the action \bear\label{AC} \tilde
S[\tilde\Phi]=-\frac{1}{2}\langle\tilde\Phi,\, \tilde
Q\tilde\Phi\rangle
-\frac13\langle\tilde\Phi,\,\tilde\Phi*\tilde\Phi\rangle, \eear
where the open string coupling constant $g_o=1$ is set for
simplicity, $\langle\cdot,\cdot\rangle$ is the BPZ inner product,
`$*$' denotes Witten's star product, and $\tilde Q$ is the new BRST
operator at the tachyon vacuum. The BRST operator $\tilde Q$ acts on
 string field $\chi$ of ghost number $gh(\chi)$ through
\bear\label{Qpsi} \tilde Q\chi = Q_B\chi + \Psi*\chi -
(-1)^{gh(\chi)}\chi * \Psi, \eear where $Q_B$ is the BRST operator
at the unstable vacuum, and $\Psi$ represents Schnabl's analytic
vacuum solution in ${\cal B}_0\Psi =0$ gauge~\cite{Schnabl:2005gv},
\bear\label{SS} \Psi\equiv \lim_{N\to\infty}\left[\sum^N_{n=0}
\psi_n^{'}-\psi_N\right] \eear with \bear\label{gss} && \psi_0 =
\frac{2}{\pi} c_1 |0\rangle,
\nn \\
&& \psi_n = \frac{2}{\pi} c_1 |0\rangle *
|n\rangle * B_1^L c_1 |0\rangle,
\qquad (n\ge 1),
\nn \\
&&\psi_0^{'}\equiv \frac{d\psi_n}{dn}|_{n=0}
= Q_B(B_1^L c_1\ket{0})= K_1^L c_1 |0\rangle + B_1^L c_0c_1|0\rangle,
\nn \\
&&\psi_n^{'}\equiv \frac{d\psi_n}{dn}
=c_1|0\rangle * K_1^L |n\rangle*B_1^Lc_1 |0\rangle,
\qquad (n\ge 1).
\eear
Here $B_1^L$ and $K_1^L$ are defined on the upper half plane(UHP)
as\footnote{We use the conventions of Ref.~\cite{Okawa:2006vm}.}
\bear\label{BK}
&&B_1^L =\int_{C_L}\frac{d\xi}{2\pi i}(1+\xi^2) b(\xi),
\nn \\
&&K_1^L =\int_{C_L}\frac{d\xi}{2\pi i}(1+\xi^2) T(\xi), \eear where
 contour $C_L$ runs counterclockwise along the unit circle with
${\rm Re}\, \xi <0$. Action (\ref{AC}) is invariant under the gauge
transformation \bear\label{gautr} \delta\tilde\Phi = \tilde
Q\tilde\Phi + \tilde\Phi*\tilde\Lambda-\tilde\Lambda*\tilde\Phi
\eear for a ghost number zero state $\tilde\Lambda$ and satisfies
the equation of motion, \bear\label{eqn1} \tilde Q\tilde\Phi +
\tilde\Phi * \tilde\Phi =0. \eear

In solving the equation of motion (\ref{eqn1}), we use a
perturbative method in some parameter $\lambda$, which is a
well-known method in OSFT. For example, the recently developed
marginally deformed solutions in OSFT used this perturbative
method~\cite{Schnabl:2007az,Kiermaier:2007ba}. The solution of
Eq.~(\ref{eqn1}) has the form,
\bear\label{sol1}
\tilde\Psi=\sum_{n=1}^{\infty} \lambda^n\tilde\phi_n
\eear
with $\tilde\phi_n$ satisfying the relations
\bear\label{Qtph1}
&&\tilde Q\tilde\phi_1 = 0,
\\ \label{Qtphn}
&&\tilde Q\tilde\phi_n = - \sum_{k=1}^{n-1} \tilde\phi_k *
\tilde\phi_{n-k}, \qquad (n\ge 2). \eear Solution (\ref{sol1})
satisfies  Eq.~(\ref{eqn1}) at each order of $\lambda$.

In this section, by explicitly solving equations (\ref{Qtph1}) and
(\ref{Qtphn}), it is shown that all perturbative solutions like
(\ref{sol1}) at the tachyon vacuum have pure gauge like
forms\footnote{ The ordinary piece $\sum_{n=0}^{\infty}
\lambda^{n+1}\psi_n^{'}$ of Schnabl's vacuum solution (\ref{SS}),
which starts from $Q_B$-exact state $\psi_0^{'}= Q_B(B_1^L
c_1\ket{0})$,  can be considered as a perturbative solution and is a
pure gauge form~\cite{Okawa:2006vm}. The solution becomes
nontrivial, however, at the special value $\lambda=1$.
For the earlier studies of
the pure gauge forms in OSFT, see
Refs.~\cite{Takahashi:2002ez,Kluson:2002av,Kluson:2003xu,
Kishimoto:2005bs}.} up to the gauge transformations. The vanishing
cohomology at the tachyon vacuum is directly connected to the form
of the solution. By introducing the homotopy operator
\bear
\label{homoperator} A=-\frac{\pi}{2}\int_1^2 dr B_1^L \ket{r},
\eear
which satisfies $\tilde Q A= {\cal I}$ with identity string
state ${\cal I}$ of star product algebra, Ellwood and Schnabl proved
that all $\tilde Q$-closed states are $\tilde Q$-exact at the
tachyon vacuum~\cite{Ellwood:2006ba}. From this fact it can be seen that
the only form of solution which satisfies Eq.~(\ref{Qtph1}) is a
$\tilde Q$-exact state
\bear\label{tph1}
\tilde \phi_1 = \tilde
Q\phi = Q_B\phi + \Psi*\phi - \phi*\Psi,
\eear
where $\phi$ is a ghost number zero state.
Unlike other backgrounds, in the
tachyon vacuum this is a unique form of solution in the $\lambda$-order of
 solution (\ref{sol1}).

From now on, we determine $\tilde\phi_n$, $(n\ge 2)$, in a given
order of $\lambda$ from the Eq.~(\ref{Qtphn}). When $n=2$,
the Eq.~(\ref{Qtphn}) is given by
\bear\label{Qtph2} \tilde
Q\tilde\phi_2 = -\tilde\phi_1*\tilde\phi_1 = -\tilde Q\phi*\tilde
Q\phi = \tilde Q\left((\tilde Q\phi)*\phi\right).
\eear
Then the following solution of the Eq.~(\ref{Qtph2}) is obtained:
\bear\label{tph2}
\tilde\phi_2 =(\tilde Q\phi)*\phi + \tilde Q\chi_2, \eear
where $\chi_2$ is a ghost number zero string state.
Up to $\lambda^2$-order, perturbative solution (\ref{sol1}) is given by
\bear\label{tilsol2}
\tilde\Psi = \lambda\tilde Q\phi +
\lambda^2\tilde\phi_2.
\eear
Inserting the Eq.~(\ref{tilsol2}) into
the action (\ref{AC}), we obtain \bear\label{AClam2}
S=\lambda^4\left[-\frac12\langle\tilde\phi_2,\, \tilde
Q\tilde\phi_2\rangle - \langle\tilde Q\phi,\, \tilde
Q\phi*\tilde\phi_2\rangle\right] + {\cal O}(\lambda^5). \eear The
terms of the $\lambda^4$-order in the action (\ref{AClam2}) are
completely determined by the truncated solution (\ref{tilsol2}).
The higher order of $\lambda$ in the action (\ref{AClam2}), however,
is not fixed by $\tilde\Psi$ given in Eq.~(\ref{tilsol2}).
As such, the valid action in this order of $\lambda$ is
the $\lambda^4$-term in Eq.~(\ref{AClam2}).
Then it can be easily found that the term $\tilde
Q\chi_2$ in Eq.~(\ref{tph2}) is a gauge degree in the action
(\ref{AClam2}) up to $\lambda^4$-order. Through the simplest
gauge choice $\tilde Q\chi_2=0$ in (\ref{tilsol2}), the following is
obtained:
\bear\label{gftilph2}
\tilde\phi_2= (\tilde Q\phi)*\phi.
\eear
Similarly, when $n=3$ in Eq.~(\ref{Qtphn}),
we have an equation \bear\label{Qtph3} \tilde
Q\tilde\phi_3=-\tilde\phi_1*\tilde\phi_2
-\tilde\phi_2*\tilde\phi_1=\tilde Q\left((\tilde
Q\phi)*\phi^2\right), \eear
where we use the notation \bear\nn
\phi^{n}\equiv \underbrace{\phi * \phi * \cdots * \phi}_{n}. \eear
From the Eq.~(\ref{Qtph3}), we obtain
\bear\label{tph3}
\tilde\phi_3=(\tilde Q\phi)*\phi^2 + \tilde Q\chi_3, \eear where
$\chi_3$ is also an arbitrary ghost number zero string state. Then
the perturbative solution (\ref{sol1}) up to the $\lambda^3$-order is
given by \bear\label{tPs3} \tilde\Psi = \lambda\tilde Q\phi +
\lambda^2(\tilde Q\phi)*\phi +\lambda^3\tilde\phi_3. \eear
Using
the same procedure as in the case of $\tilde\phi_2$, we can fix
the gauge by choosing $\tilde Q\chi_3=0$ from the action in the
$\lambda^5$-order. Then the gauge fixed $\tilde\phi_3$ will be given by
\bear\label{tph3-1} \tilde\phi_3 = (\tilde Q\phi)*\phi^2. \eear
By repeating the aforementioned procedures, the gauge fixed
$\tilde\phi_n$ can be obtained from $\lambda^{n+2}$-term in OSFT action
(\ref{AC}): \bear\label{tphn-1} \tilde\phi_n = (\tilde
Q\phi)*\phi^{n-1}. \eear As a result, perturbative solution
(\ref{sol1}) is represented as a pure gauge solution\footnote{By
using $\tilde Q A={\cal I}$ at the tachyon vacuum, the following pure
gauge solution can also be constructed:
\bear\label{anopure}
\tilde\Psi= \sum_{n=1}^\infty\lambda^n(\tilde Q\phi)*(A*\tilde
Q\phi)^{n-1}. \eear This solution can be constructed only at the tachyon
vacuum. The solution (\ref{anopure}) can also be reduced to the
simplest solution (\ref{sol2}) by gauge fixing it from the relation
$A*\tilde Q\phi = \phi -\tilde Q (A \phi)$.}, \bear\label{sol2}
\tilde\Psi = \sum_{n=1}^{\infty} \lambda^n (\tilde Q\phi) *
\phi^{n-1} =\tilde Q\phi *\frac{\lambda}{1-\lambda\phi} =
e^{-\tilde\Lambda}* \left(\tilde Q e^{\tilde\Lambda}\right), \eear
where the ghost number zero string state $\tilde\Lambda$ is given by
\bear\nn \tilde\Lambda = -\ln (1-\lambda\phi) = \sum_{n=1}^{\infty}
\frac{\lambda^n}{n}\,\phi^n. \eear
For any pure gauge form of
solution $\tilde\Psi=\tilde U\tilde Q\tilde U^{-1}$ with a ghost
number 1 string state $\tilde U$, we can find a perturbative type
solution by setting $\lambda\phi= {\cal I}- \tilde U$ at the tachyon
vacuum. A similar pure gauge form in terms of BRST operator $Q_B$
for Schnabl's vacuum solution was found by Okawa~\cite{Okawa:2006vm}.

Since the action (\ref{AC}) is invariant under the infinitesimal
(small) gauge transformation (\ref{gautr}), it is also invariant
under the gauge function $e^{\tilde\Lambda}$ in Eq.~(\ref{sol2})
which is generated by small gauge transformations. In this case,
where $e^{\tilde\Lambda}$ can be deformed to the identity string
state ${\cal I}$, the pure gauge solution (\ref{sol2}) has no
physical meaning. In some cases, however, $e^{\tilde\Lambda}$ state
cannot be deformed continuously into the identity string state at the
tachyon vacuum. Then the action is not invariant for the pure gauge
solution (\ref{sol2})(i.e., $\tilde S[e^{-\tilde\Lambda}\tilde Q
e^{\tilde\Lambda}] \ne 0$).
In the gauge theory, this type of gauge
transformation is called {\it large gauge transformation}. In
this case, the pure gauge solution has a nontrivial physical
meaning\footnote{In the construction of marginal deformation for
photon around the unstable vacuum, the pure gauge solution was used
by Fuchs et al.~\cite{Fuchs:2007yy,Fuchs:2007gw}. In the paper the
non-normalizable states were used  and the counterterms were added
to obtain some meaningful results which are independent of the
non-normalizable states.}. In the subsection 3.2, an explicit
form of $\phi$, which corresponds to the large gauge
transformation, will be introduced.

\section{Marginal Deformations}

The string state $\tilde\phi_n$ in Eq.~(\ref{tphn-1}) is a wedge
state~\cite{Rastelli:2000iu} with operator insertions on the
worldsheet boundary. If string state $\phi$ in the construction
of solution (\ref{sol2}), which is made from some operator
insertion into the SL(2,R) vacuum $\ket{0}$, is well-defined,
$\tilde\phi_n$ will not cause any divergence in the calculations of
BPZ inner products since the separations among the boundary
insertions do not go to zero by construction. As we discussed in the
previous section, the only constraint for $\phi$ is its ghost number
(i.e., $gh(\phi)=0$). Since the solution herein, however,
around the tachyon vacuum, is the pure gauge form,
choosing the physically acceptable matter operator
will become nontrivial.

An important and well-known solution in OSFT is the marginally
deformed solution~\cite{Schnabl:2007az,Kiermaier:2007ba}. As an
application of the construction, we consider matter operators which
give exactly marginal deformations in BCFT~\cite{Callan:1994ub}.
These operators are of particular interest in the study of tachyon
condensation and the cohomology class of BRST operator $Q_B$ at the
unstable vacuum.

We can also construct a pure gauge solution around the unstable
vacuum by replacing the BRST operator $\tilde Q$ with $Q_B$ in the
Eq.~(\ref{sol2}), apart form the issue of the gauge invariance
 of pure gauge form.
However, we cannot obtain the marginal deformation by the pure gauge
solution around the unstable vacuum, except for the cases in
Refs.~\cite{Fuchs:2007yy,Fuchs:2007gw} which have special
prescriptions. The reason is following: The first term in the pure
gauge solution for the BRST operator $Q_B$ is \bear\nn \lambda
Q_B\phi, \eear where $\phi$ is a ghost number zero string state and
include an exactly marginal operator V. To extract the contribution
to the marginal deformation $c_1V(0)\ket{0}$, we can use a test
state $c_0c_1\tilde V(0)\ket{0}$ with a primary operator $\tilde V$.
But the contribution of $\lambda Q_B\phi$ to $c_1V(0)\ket{0}$
vanishes always since \bear\nn \langle c_0 c_1\tilde V,\,
Q_B\phi\rangle = -\langle Q_B(c_0 c_1\tilde V),\,\phi\rangle=0.
\eear Therefore, we cannot obtain the marginally deformed solution
from the pure gauge solution around the unstable vacuum without
special prescriptions.

\subsection{Marginally deformed operators}

Let us consider an exactly marginal operator called $V$ with
conformal dimension one in the construction of ghost number zero
string state $\phi$ in Eq.~(\ref{sol2}). After some investigations
of the construction of $\phi$, we found that the most plausible and
simple choice for $\phi$ is\footnote{For the special choice of
$\phi=V(0)B_1^L c_1\ket{0}$, the pure gauge solution (\ref{anopure})
is the same as the solution (\ref{sol2}) since $A*\tilde Q\phi=
A*Q_B\phi + A*\Psi*\phi-A*\phi*\Psi=\phi$, where we used the facts,
$A*\phi=0$, $Q_B A = {\cal I} -\ket{0}$, and $A*\Psi=
B_1^Lc_1\ket{0}$.} \bear\label{chphi} \phi=V(0) B_1^L c_1 \ket{0},
\eear where $B_1^L$ was defined in Eq.~(\ref{BK}). Due to the
properties of $B_1^L$, the solution is significantly simplified.
The other ghost number zero states, for instance,
$\phi=V(0)\ket{0}$ or $\phi=V(0)B_1^Lc_0\ket{0}$, can also be considered.
These
cases, however, do not give marginal deformations since the term
$V(0)c_1\ket{0}$, which corresponds to the marginal deformation in
the $\lambda$-order in the resulting solution $\tilde\Psi$ in
Eq.~(\ref{sol2}), vanishes. In these reasons, we suggest a special
choice for $\phi$ given in Eq.~(\ref{chphi}), and we examined this
fact for the rolling tachyon marginal operator $V=e^{\pm X^0}$, which
has a nontrivial physical meaning, as will be seen in the next
subsection. For the general marginal operators, further investigation
is needed.

In what follows in this section, we restrict our interest to the
case of $\phi$ given in Eq.~(\ref{chphi}). As was discussed
about our pure gauge solution in section 2, we have to choose the
matter operator $V$ in the gauge function, \bear\label{gauftn}
e^{\tilde\Lambda} = \frac{1}{1-\lambda\phi}= \frac{1}{1-\lambda V(0)
B_1^L c_1\ket{0}}, \eear which gives nontrivial physical meanings.

Under the assumption that a matter operator $V$ makes the solution
(\ref{sol2}) nontrivial, we write down the explicit form of
$\tilde\phi_n$ in Eq.~(\ref{tphn-1}) as
\bear\label{tphn2}
\tilde\phi_n&=&\left(V(0) c_0\ket{0} + \partial V(0) c_1\ket{0}
+ V(0) c_1K_1^L\ket{0}\right)*J^{n-2}*V(0)B_1^Lc_1\ket{0}
\nn \\
&& + \Psi*J^{n-1}*V(0)B_1^Lc_1\ket{0}
\nn \\
&&-V(0)B_1^Lc_1\ket{0}*\Psi*J^{n-2}*V(0)B_1^Lc_1\ket{0},
\eear
where $J\equiv V(0)\ket{0}$, and the following relations are
used:
\bear\label{QBphn}
&&Q_B\phi = V(0) c_0\ket{0} + \partial V(0) c_1\ket{0}
+ V(0) c_1K_1^L\ket{0},
\nn\\
&&\phi^n=J^{n-1} * V(0)B_1^Lc_1\ket{0}.
\eear
In obtaining the expression $\phi^n$ in Eq.~(\ref{QBphn}), we
used the relations,
\bear
&&B_1^L\phi_1*\phi_2 = B_1\phi_1*\phi_2
+ (-1)^{gh(\phi_1)}\phi_1*B_1^L\phi_2,
\nn \\
&&\{B_1,\,c_1\}=1,\quad B_1\ket{0}=0,\quad (B_1^L)^2=0.
\nn\eear

Recently, marginally deformed exact solutions around the unstable
vacuum are obtained in bosonic
string~\cite{Schnabl:2007az,Kiermaier:2007ba,
Fuchs:2007yy,Okawa:2007it,Kishimoto:2007bb,Kiermaier:2007vu,Lee:2007ns}
and superstring~\cite{Erler:2007rh,Okawa:2007ri,Kishimoto:2007bb,
Fuchs:2007gw,Kiermaier:2007ki} field theories. If the operator $V$
has a regular OPE with itself, the solutions are well-defined.
When the OPE of $V$, however, is singular, divergencies arise in the
solutions, and one needs to add counterterms to regularize it at
each order of $\lambda$~\cite{Kiermaier:2007ba} or renormalize the
operator $V$~\cite{Kiermaier:2007vu,Kiermaier:2007ki}. In the
construction of $\tilde\phi_n$ in Eq.~(\ref{tphn2}), there is no
divergence even if the case for the operator $V$ with
singular OPE would be considered, as explained earlier.

\subsection{Rolling tachyon: Late time behaviors of D-brane decay}

As a very important application of the solution (\ref{sol2}), we
consider the rolling tachyon vertex operator,
$V(y)=e^{\pm\frac{1}{\sqrt{\alpha'}} X^0(y)}$,
where $y$ is the boundary coordinate on UHP.
In $\alpha'=1$ units, $V$ is the dimension one primary operator.
Since we are considering the string field theory around the tachyon
vacuum, we choose
\bear\label{taV}
V(y) = e^{-X^0(y)}
\eear
for definiteness. In physical point of view,
the deformation of tachyon field
$e^{-X^0(0)}c_1\ket{0}$ in the $\lambda$-order represents the facts
that the system will reach to the tachyon vacuum in the far future.

By inserting rolling tachyon vertex operator (\ref{taV}) into
the Eq.~(\ref{chphi}), we obtain
\bear\label{chphi2}
\phi= e^{-X^0(0)} B_1^L c_1\ket{0}.
\eear
Then the gauge function $e^{\tilde\Lambda}$ is given by
\bear\label{gauftn2}
e^{\tilde\Lambda}= \frac{1}{1-\lambda e^{-X^0(0)} B_1^L c_1\ket{0}}.
\eear
The above gauge function is an example that generates the so-called
{\it large gauge transformation} corresponding to nontrivial physics.
The reason for this is the following:
One cannot deform the gauge function
$e^{\tilde\Lambda}$ in Eq.~(\ref{gauftn2}) to identity string
state ${\cal I}$ since we can always rescale the gauge parameter
$\lambda$ by the time $x^0$ (zero mode of $X^0$) translation.

From the expression for $\tilde\phi_n$ given in Eq.~(\ref{tphn2})
we obtain the string field solution
corresponding to rolling tachyon marginal deformation,
\bear\label{sol3}
\tilde\Psi&=& \sum_{n=1}^\infty\lambda^n\tilde\phi_n
=\sum_{n=1}^\infty \lambda^n(A_n + B_n + C_n)
\nn \\
&=&\sum_{n=1}^\infty\lambda^n\left(\beta_n e^{-n X^0(0)}c_1\ket{0}
+\cdots \right),
\eear
where $\beta_n$ is the coefficient of the tachyon profile and
\bear\label{ABC}
A_n
&=& (Q_B\phi)*\phi^{n-1}
\nn \\
&=& \left(e^{-X^0(0)}c_0\ket{0} + (\partial e^{-X^0(0)})c_1\ket{0}
+e^{-X^0(0)}c_1K_1^L\ket{0}\right)*J^{n-2}*e^{-X^0(0)}B_1^L c_1\ket{0},
\nn \\
B_n&=& \Psi*\phi^n
\nn \\
&=&\Psi*J^{n-1}*e^{-X^0(0)}B_1^Lc_1\ket{0},
\nn \\
C_n&=&-\phi*\Psi*\phi^{n-1}
\nn \\
&=&-e^{-X^0(0)}B_1^L c_1\ket{0} *\Psi*J^{n-2}*
e^{-X^0(0)}B_1^Lc_1\ket{0}
\eear
with $J=e^{-X^0(0)}\ket{0}$.
In the third step of the Eq.~(\ref{sol3}), the tachyon component was
separated, and $\cdots$ indicates the higher level fields.

By using a test string state
\bear\label{teststate}
\chi_n = e^{nX^0(0)} c_0c_1\ket{0},
\eear
the coefficient $\beta_n$ in Eq.~(\ref{sol3}) can be extracted by
\bear\label{betn}
\beta_n=\langle \chi_n,\,\tilde\phi_n\rangle,
\eear
where we omit volume factor $(2\pi)^D\delta^D(0)$ for the $D$
spacetime dimensions arising from the BPZ inner products
$\langle\cdot,\cdot\rangle$ for simplicity.
For the conventions of the matter and ghost correlation functions,
see Appendix. Then the exact expression for $\beta_n$ is given by
\bear\label{betn2}
\beta_n = \beta_n^A + \beta_n^B + \beta_n^C,
\eear
where
\bear\label{betnABC}
\beta_n^A&\equiv& \langle\chi_n,\, A_n\rangle
\nn \\
&=&\frac{1}{2}
\left(\frac{2}{\pi}\right)^{n^2+n}
\frac{\partial}{\partial x}\left[ a^{n^2+n-1}\left\{
\left(x-\frac{1}{2a}\sin (2ax)\right)\sin^2a + \left(1-\frac{1}{2a}
\sin(2a)\right)\sin^2(ax)\right\}\right.
\nn \\
&&\hskip 2.9cm
\times\frac{\prod_{j=2}^n\sin^2(a(x-j))\prod_{2\le k < m}^n
\sin^2(a(k-m))}{\sin^{2n}(ax)\prod_{i=2}^n\sin^{2n}(ai)}\Bigg]_{x=1}
\nn \\
&&+\,\left(\frac{2}{\pi}\right)^{n^2+n}
\frac{\partial}{\partial m}\left[
 b^{n^2+n-1} \sin^2 b
\left(1-\frac{1}{2b} \sin(2b)\right)\right.
\nn \\
&&\hskip 3.1cm \times \frac{\prod_{j=0}^{n-2}\sin^2(b(m+j-1))
\prod_{0\le k <l}^{n-2}\sin^2(b(k-l))}{
\sin^{2n}b\prod_{i=0}^{n-2}\sin^{2n}(b(m+i))}\Bigg]_{m=2},
\nn \\
\beta_n^B&\equiv& \langle\chi_n,\, B_n\rangle
\nn \\
&=&\left(\frac{2}{\pi}\right)^{n^2+n}
\sum_{m=0}^\infty\frac{\partial}{\partial m}\left[
 c^{n^2+n-1} \sin^2 c\left(
1-\frac{1}{2c}\sin(2c)\right)
\frac{\prod_{2\le j<k}^{n+1}\sin^2(c(j-k))}{
\prod_{i=2}^{n+1}\sin^{2n}(c(m+i))}\right],
\nn \\
\beta_n^C&\equiv&\langle\chi_n,\, C_n\rangle
\nn \\
&=&-2\left(\frac{2}{\pi}\right)^{n^2+n}
\sum_{m=0}^\infty\frac{\partial}{\partial m}\left[
c^{n^2+n-1}\cos c \sin^2 c
\left(\cos c -\frac{1}{c}\sin c\cos(2c)\right)\right.
\nn \\
&&\hskip 4cm \times\frac{\prod_{j=3}^{n+1}\sin^2(c(m+j-1))
\prod_{3\le k<l}^{n+1} \sin^2(c(k-l))}{ \sin^{2n}
c\prod_{i=3}^{n+1}\sin^{2n}(c(m+i))}\Bigg] \eear with \bear\nn
a=\frac{\pi}{n+1},\qquad b=\frac{\pi}{m+n-1},\qquad
c=\frac{\pi}{m+n+2}. \eear In the calculations of $\beta_n^B$ and
$\beta_n^C$, we have to use the Schnabl's vacuum solution $\Psi$
which is composed of the ordinary piece
$\sum_{n=0}^{\infty}\psi_n^{'}$ and the phantom piece $-\psi_\infty$.
The ordinary piece has nontrivial contributions to $\beta_n$ and
makes a convergent series. However, the phantom piece has no
contribution to $\beta_n$ since the coefficients of $\psi_N$ in
$L_0$-level truncation (level zero in this case) go to zero as ${\cal
O}(N^{-3})$ for a large $N$. The detailed calculations for BPZ-inner
products of $\beta_n^A$, $\beta_n^B$, and $\beta_n^C$ are given in
Appendix.

\begin{figure}\label{Fig1}
\centerline{\includegraphics[width=115mm]{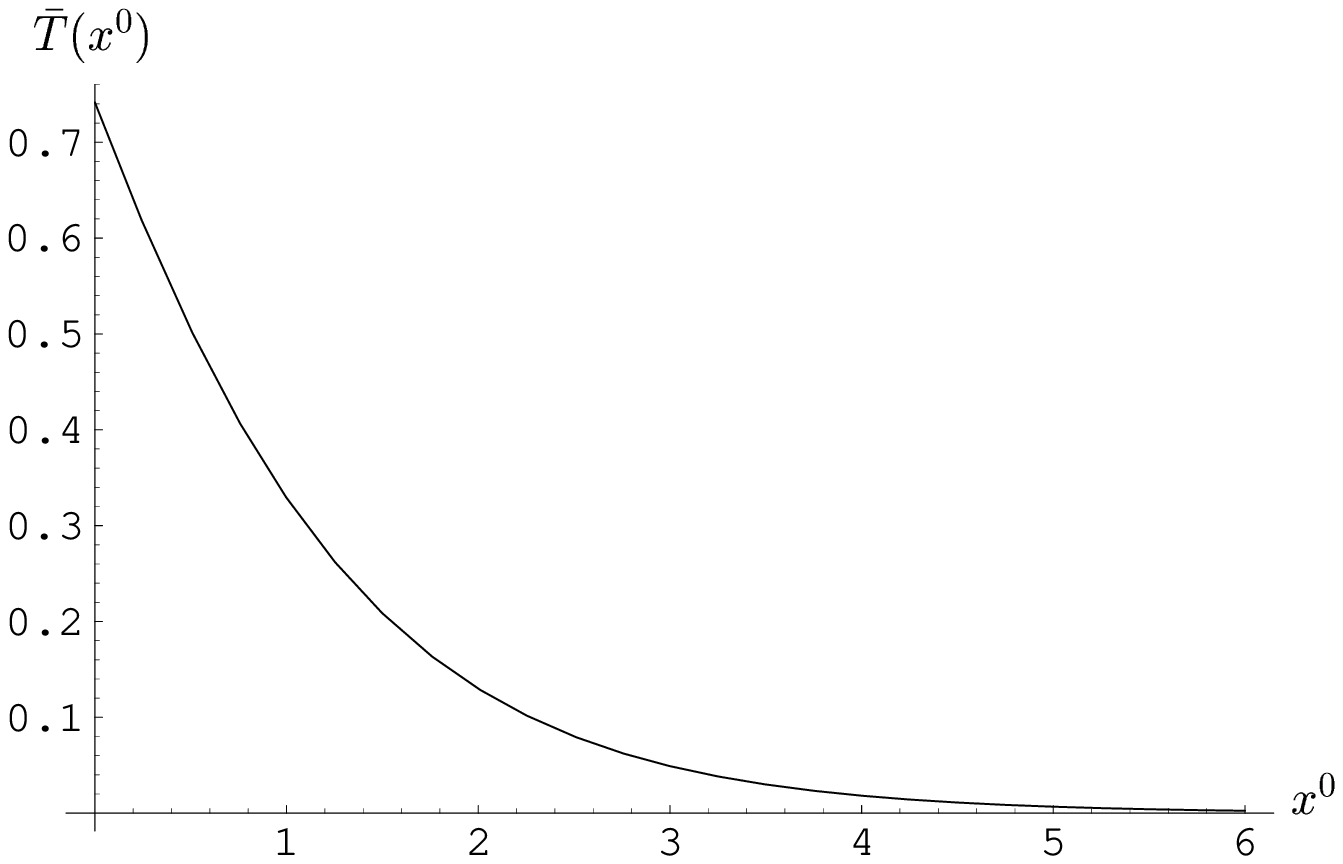}}
\caption{{\small Graph of $\bar T(x^0)$.}}
\end{figure}
\begin{table}\label{cvtst}
\begin{center}\def\st{\vrule height 2.7ex width 0ex}
\begin{tabular}{|c|c|c|c|c|c|c|} \hline
 & $n$=1 & $n$=2 & $n$=3 & $n$=4 & $n$=5& $n$=6
\st\\[0ex]
\hline $\bar\beta_n^A$ &0.0 & 3.9161 & 0.096432 & 9.0328$\times
10^{-5}$ & 2.4981$\times 10^{-9}$ &1.4941$\times 10^{-15}$
\st\\[0ex]
\hline $\bar\beta_n^B$ &8.7380  & 0.35979 &0.00080278 &
5.9455$\times 10^{-8}$&9.7343$\times 10^{-14}$ &2.5567$\times
10^{-21}$
\st\\[0ex]
\hline $\bar\beta_n^C$ & -7.7380& -3.9711 &-0.050663  &
-1.9498$\times 10^{-5}$ &-1.6748$\times 10^{-10}$  &-2.2996$\times
10^{-17}$
\st\\[0ex]
\hline\hline $\bar\beta_n$ &1.0 &0.30483 &0.046572  & 7.0889$\times
10^{-5}$
 & 2.3307$\times 10^{-9}$&1.4711$\times 10^{-15}$
\st\\[0ex]
\hline
\end{tabular}
\end{center}
\caption{{\small Several few coefficients for $\bar\beta_n^A$,
$\bar\beta_n^B$, $\bar\beta_n^C$,  and $\bar\beta_n$. }}
\end{table}
The numerical results for the first few $\beta_n$ are
\bear\label{betn3}
&&\beta_1= -0.042740,\quad \beta_2 = - 0.013018, \quad
\beta_3 = -0.0019905,
\nn \\
&&\beta_4 = -3.0298\times 10^{-6},\quad \beta_5= -9.9612\times
10^{-11},\quad \beta_6= -6.2872\times 10^{-17}. \eear
We can easily
obtain the higher coefficients of $\beta_n$ by adjusting the number
of significant digits to increase numerical precision in the computer
program. In the convention herein, the tachyon potential is unbounded
from below at $T\to +\infty$. As such, the value of the
tachyon field at the unstable vacuum is greater than
that at the tachyon vacuum. Since the deformation at the true vacuum
is being considered, the rolling
tachyon deformation, which describes the decay of the unstable D-brane
from the unstable vacuum to the tachyon vacuum, is positive, i.e.,
$T>0$. As we have shown in Eq.~(\ref{betn3}), all coefficients
$\beta_n$ are negative. Therefore, the physical solution for the
tachyon profile in Eq.~(\ref{tasol}) corresponds to $\lambda <0$.
After rescaling by translation in the time direction, we can set
$\lambda=-1$ in Eq.~(\ref{sol3}).

If we normalize $\beta_n$ by using $\beta_1$ for convenience,
 the resulting tachyon profile in Eq.~(\ref{sol3}) is given by
\bear\label{tasol}
\bar T(X^0)&\equiv& \frac{T(X^0)}{|\beta_1|}
\nn \\
&=&  e^{-X^0} + \sum_{n=2}^\infty (-1)^{n+1}\bar\beta_n
e^{-n X^0} \nn \\
&=&  e^{-X^0} - 0.3048\, e^{-2 X^0} +
0.04657\, e^{-3 X^0} -7.089 \times 10^{-5}\, e^{-4 X^0}
\nn \\
&&+ 2.331\times 10^{-9}\, e^{-5 X^0} -1.471 \times 10^{-15}\,  e^{-6
X^0} + \cdots, \eear where $\bar\beta_n \equiv\beta_n/\beta_1$. The
behaviors of the rescaled tachyon field $\bar T(x^0)$ are plotted in
Fig.1. To obtain physical intuitions for the roles of BRST operator
$Q_B$ and the vacuum solution $\Psi$ in the tachyon profile, we
summarize the first few normalized coefficients, $\bar\beta_n^A
\equiv\beta_n^A/\beta_1$, $\bar\beta_n^B \equiv\beta_n^B/\beta_1$,
and $\bar\beta_n^C \equiv\beta_n^C/\beta_1$ in Table 1. As can be
seen in Table 1, in the late time behaviors of the tachyon profile
(encoded in $\bar\beta_1$) the roles of tachyon vacuum solution
$\Psi$ are crucial in the rolling tachyon deformation, and there is
no contribution from $\bar\beta_1^A$ which comes from the BRST
charge $Q_B$. From the next order of tachyon coefficients $\beta_n$,
($n\ge 2$), the contributions of tachyon vacuum solution (encoded in
$\bar\beta_n^B$ and $\bar\beta_n^C$) rapidly decrease and those of
$Q_B$-term become dominant.

Tachyon field $T(x^0)$ decreases monotonically and approaches zero
at $x^0\to\infty$, as can be seen in Fig.1. The system approaches the
tachyon vacuum without oscillating behaviors at a late time of the
D-brane decay.
A similar behavior of the tachyon profile at a late time was
reported in Ref.~\cite{Ellwood:2007xr} by neglecting the
contributions of matter correlators.
However, our result suggests
that in the far past, the tachyon profile had wild oscillating
behaviors.  Therefore, it seems that our solution connects the
wild oscillations to the tachyon vacuum.
Since the tachyon field, however, is not gauge invariant,
it is difficult to identify the oscillating behaviors in the far
past in this paper as those of the marginally deformed
rolling tachyon around the unstable vacuum, which were obtained
in Refs.~\cite{Schnabl:2007az,Kiermaier:2007ba}.

\section{Conclusion}

In this work, we investigated the analytic solutions around the
tachyon vacuum in OSFT. Based on the fact that all $\tilde Q$-closed
states at the tachyon vacuum are $\tilde Q$-exact due to the
existence of homotopy operator $A$~\cite{Ellwood:2006ba}, we
construct a pure gauge solution from the perturbative type of
solution (\ref{sol1}). This situation at the tachyon vacuum in OSFT
is reminiscent of that at the spatial infinity in gauge theory,
such as the four-dimensional Euclidean Yang-Mills theory
with an instanton solution.
Like the gauge theory, the pure gauge solutions, which
correspond to the {\it large gauge transformations}, are physically
nontrivial.

As an application of the construction herein of a
pure gauge solution around
the tachyon vacuum, we considered the marginally deformed rolling
tachyon vertex operator $V=e^{-X^0}$, in which the system stays at
the tachyon vacuum in the far future. After some investigations, we
found that a special choice of the ghost number zero string state
$\phi$ in Eq.~(\ref{chphi2}) and corresponding gauge function
$e^{\tilde\Lambda}$ given in Eq.~(\ref{gauftn2}) make the pure
gauge solution nontrivial for the rolling tachyon vertex operator.
In this special choice, the pure gauge solution corresponds to the
 large gauge transformation. Then the gauge function
$e^{\tilde\Lambda}$ cannot be deformed to the identity string state
${\cal I}$ since gauge parameter $\lambda$ represents a
translation along the time direction. We found that under the choice
of $\phi$ in Eq.~(\ref{chphi2}), we could make the marginal
deformation due to the nontrivial roles of the analytic tachyon
vacuum solution $\Psi$ in Eq.~(\ref{SS}). We explicitly obtained the
tachyon profile, which is believed to describe the behaviors of
D-brane decay at late times. The coefficients $\beta_n$ represent
exponentially decreasing behaviors, which are similar to those of
tachyon coefficients at the unstable
vacuum~\cite{Schnabl:2007az,Kiermaier:2007ba}. According to the
results of this study,
the tachyon profile decrease monotonically and approaches
the tachyon vacuum asymptotically at a late time of the D-brane decay.
These behaviors of the tachyon profile around the tachyon vacuum
were also obtained in Ref.~\cite{Ellwood:2007xr} by neglecting the
roles of matter correlators. Our results, however, show that
there were wild oscillations in the far past. Therefore, our solution
seems to connect the wild oscillations of tachyon profile to the
tachyon vacuum.

By using the remarkable property of OSFT, we can relate the
marginally deformed rolling tachyon around the unstable vacuum and
our solution around the tachyon vacuum.
If $\tilde\Psi_u$ is the rolling tachyon
solution~\cite{Schnabl:2007az,Kiermaier:2007ba}
satisfying $Q_B\tilde\Psi_u + \tilde\Psi_u *\tilde\Psi_u=0$ around the
unstable vacuum, the string field $(-\Psi+\tilde\Psi_u)$ is the
solution around the tachyon vacuum since it satisfies the equation
of motion,
\begin{eqnarray}
&&\tilde Q(-\Psi + \tilde \Psi_u)+(-\Psi + \tilde \Psi_u) * (-\Psi +
\tilde \Psi_u) \nonumber \\ && = Q_B\tilde\Psi_u +
\tilde\Psi_u*\tilde\Psi_u=0,\nonumber
\end{eqnarray}
where we used the fact that $Q_B\Psi + \Psi*\Psi=0$. Though $(-\Psi
+ \tilde\Psi_u)$ and our solution $\tilde\Psi$ satisfy the same
equation of motion and use the same rolling tachyon vertex operator,
the two solutions are gauge inequivalent since the action values for
the two are different due to the presence of Schnabl's solution in
the former case. Here we used the fact that the BPZ inner products
including $\tilde\Psi_u$ or $\tilde\Psi$ in the calculation of the
actions are trivially zero from the momentum conservation in the
products. Therefore to find the relation between $\tilde\Psi_u$ and
$\tilde\Psi$, we have to obtain the late time behaviors of
$\tilde\Psi_u$ around the tachyon vacuum. However, it was not known
yet.

If the two rolling tachyon marginal solutions around the unstable
vacuum and the solution presented herein
describe the same physical situation, although further investigation
in this direction is needed, the result of this study suggests
the following possibilities. The first possibility is that
the puzzling oscillating behaviors can be eliminated
through the resummation of the series form
of the marginal solutions, if such is possible.
Actually, this possibility
is not very promising since the oscillating behaviors were examined by
using various methods in
literatures~\cite{Moeller:2002vx,Fujita:2003ex,Coletti:2005zj,
Forini:2006tn}. Even in the analytic
solution~\cite{Schnabl:2007az,Kiermaier:2007ba}, the oscillating
behaviors were almost confirmed from the behaviors of the coefficients
of the tachyon profile. The results presented in this paper
 also support
the oscillating behaviors before the system approaches the tachyon
vacuum. The other possibility is that by using a time dependent
gauge transformation, the rolling tachyon solution, which connects
the unstable vacuum to the tachyon vacuum without the
oscillating behaviors, as suggested in
literatures~\cite{Coletti:2005zj,Ellwood:2007xr}, can be obtained.
Although it is
difficult to give concluding remarks based on the results of this
study, we
think that our results can, hopefully, shed some light on the puzzle
of rolling tachyon solution (tachyon matter problem) in OSFT, since
we fixed the behaviors of the tachyon profile at the late time of
D-brane decay. Since the tachyon field alone, however, is not gauge
invariant, the physical meanings of the tachyon profile, which were
obtained through marginal deformation at the unstable vacuum, and our
result at the tachyon vacuum are not clear. Further investigation
in this direction is needed.

As we have seen in Refs.~\cite{Fuchs:2007yy,Fuchs:2007gw} or
in Schnabl's vacuum solution\footnote{For the breakdown of gauge
symmetry in the ordinary piece when $\lambda=1$,
see the section 5 of Ref.~\cite{Ellwood:2006ba}.},
 pure gauge solutions are useful in obtaining physically
nontrivial solutions. Of course, to obtain meaningful results,
gauge degree must be avoided.
In this sense, the construction of pure gauge
solution at the tachyon vacuum herein can be used to obtain
meaningful solutions,
for instance, marginal solutions or soliton solutions,
through various methods.
Extension of our method to supersymmetric string field theory
is also an interesting subject.

\section*{Acknowledgements}
I am grateful to Stefano Kovacs for very useful
discussions and comments.
This work was supported by SFI Research Frontiers Programme in Ireland,
the Korea Research Foundation Grant funded by the Korean
Government(MOEHRD)(KRF-2006-352-C00010), the Astrophysical Research
Center for the Structure and Evolution of the Cosmos (ARCSEC)), and
grant No. R01-2006-000-10965-0 from the Basic Research Program
through the Korea Science $\&$ Engineering Foundation.

\appendix

\section{Conventions and Calculations for $\beta_n^A$, $\beta_n^B$,
and $\beta_n^C$}

We calculate the correlators, which come from BPZ-inner products,
in semi-infinite cylinder(SIC) (frequently called as sliver frame) with
coordinate $z$.
The coordinate $z$ has the relation
\bear\label{zxi}
z=f(\xi)=\frac{2}{\pi} \tan^{-1}\xi
\eear
with coordinate $\xi$ in UHP.
We basically follow the notations used in
literatures~\cite{Okawa:2006vm,Okawa:2006sn,Kiermaier:2007ba}.

Using appropriate mappings $f_i(\xi)$ from UHP to SIC,
we can obtain the following relation between BPZ-inner product and
correlation function in conformal field theory
for general type of BPZ-inner product,
\bear\label{pppn}
&&\langle\phi_0,\, \phi_1*\phi_2*\cdots\phi_n\rangle
\nn \\
&&=\langle f_0\circ\phi_0(0)f_1\circ\phi(0) \cdots
f_n\circ\phi_n(0)\rangle_{{\cal W}_1},
\eear
where $\langle\cdots\rangle_{{\cal W}_n}$ denotes a correlation
function on SIC with ${\cal W}_n$ called as wedge state surface with
circumference,
\bear
-\frac12 (1+n) \le {\rm Re} (z) \le \frac{1}{2} (1 + n),
\eear
$\phi_i$ denotes a generic state in the Fock space and
$\phi_i(0)$ represents the corresponding operator having the
relation $\ket{\phi_i}=\phi_i(0)\ket{0}$, and
\bear
f_j(\xi) = \frac{2j}{n+1} + \frac{4}{(n+1)\pi}\tan^{-1}\xi,
\quad (0\le j \le n).
\eear
Then under the assumption that $\phi_i(0)$ is a primary operator
with conformal dimension $h_i$ for simplicity, we obtain
\bear\label{BPZcor}
&&\langle\phi_0,\, \phi_1*\phi_2*\cdots\phi_n\rangle
\nn \\
&& =\left(\frac{4}{(n+1)\pi}\right)^{\sum_{i=0}^nh_i}
\Big\langle\phi_0(0) \phi_1\left(\frac{2}{n+1}\right) \cdots
\phi_j\left(\frac{2 j}{n+1}\right)\cdots
\phi_n\left(\frac{2 n}{n+1}\right)\Big\rangle_{{\cal W}_1}
\nn \\
&&=\left(\frac{2}{\pi}\right)^{\sum_{i=0}^nh_i}
\langle\phi_0(0)\phi_1(1)\cdots\phi_n(n)\rangle_{{\cal W}_{n}},
\eear
where in the second step we rescaled the coordinate,
$z\to \frac{n+1}{2}\, z$.

\subsection{Ghost and matter correlators}

We use the following convention for the ghost correlator on UHP,
\bear\label{COR}
\langle c(\xi_1) c(\xi_2) c(\xi_3)\rangle_{UHP} =
(\xi_1-\xi_2)(\xi_1-\xi_3)(\xi_2-\xi_3).
\eear
From the conformal transformation from UHP to SIC, we obtain
the ghost correlator on the wedge state surface ${\cal W}_\alpha$,
\bear\label{COR3}
\langle c(z_1) c(z_2) c(z_3)\rangle_{{\cal
W}_\alpha,\,g} = \left(\frac{1+\alpha}{\pi}\right)^3 \sin s_{12}
\sin s_{13} \sin s_{23}
\eear
with definition $s_{ij}\equiv \frac{\pi( z_i- z_j)}{1+\alpha}$,
where $\langle\cdot\rangle_{{\cal W}_\alpha,\, g}$ denotes the
ghost correlator.
Using the relations $\{{\cal B}_0+{\cal B}_0^\star,\, c(z)\}=z$ and
$\{{\cal B},\, c(z)\}=1$, we obtain a very useful formula in the
calculations of ghost correlators,
\bear\label{Bcccc2}
\langle {\cal B} c(z_1) c(z_2) c(z_3) c(z_4)
\rangle_{{\cal W}_\alpha,\, g}
&=&\frac{(1+\alpha)^2}{\pi^3}\left[
- z_1 \sin s_{23} \sin s_{24} \sin s_{34}
\right. \nn \\
&&\left.
\hskip 1.7cm +  z_2 \sin s_{13} \sin s_{14} \sin s_{34}
\right. \nn \\
&&\left.
\hskip 1.7cm- z_3 \sin s_{12} \sin s_{14} \sin s_{24}
\right. \nn \\
&& \left.
\hskip 1.7cm+  z_4 \sin s_{12} \sin s_{13} \sin s_{23}\right],
\eear
where ${\cal B}_0$ is the zero mode of the $b$ ghost in the $z$
coordinate, ${\cal B}_0^\star$ is its BPZ conjugate, and
\bear\label{BBB}
{\cal B} = \int \frac{dz}{2\pi i}\, b(z)=
\frac{\pi}{2}\,f\circ B_1^L.
\eear
When ${\cal B}$ is located between two operators
at $t_1$ and $t_2$  with $\frac12 <t_1 <t_2$, the contour of the
integral can be taken to be $-V_\alpha^+$ with $2t_1 -1 <\alpha <2t_2-1$.
Here the oriented straight lines $V^{\pm}_\alpha$ in SIC is defined as
\bear
&&V_\alpha^{\pm} = \left\{ z\Big |\, \mbox{Re} (z) =
\pm\frac12 (1 +\alpha)\right\}, \nn \\
&& \mbox{orientation} \,:\, \pm\frac12 (1 +\alpha) -i \infty
\,\longrightarrow\, \pm\frac12 (1 +\alpha) + i\infty.
\eear

On the other hand, for the matter correlators, we use the two
point function
\bear\nn
\langle X^\mu (\xi) X^\nu (\xi')\rangle_{UHP} = - 2
\eta^{\mu\nu} \ln |\xi - \xi'|.
\eear
Then the general $n$-point correlator on UHP is given by
\bear\nn
&&\langle e^{ik_1\cdot X(\xi_1)}
e^{ik_2\cdot X(\xi_2)} \cdots  e^{ik_n\cdot X(\xi_n)}\rangle_{UHP}
\nn \\
&&= (2\pi)^D\delta^D (k_1 + k_2 + \cdots+ k_n)\prod_{1\le i<j}^n
|\xi_i-\xi_j|^{2 k_i\cdot k_j}
\eear
Using the conformal transformation from UHP to SIC, we obtain
the matter correlator on SIC,
\bear\label{npc2}
&&\langle
e^{ik_1\cdot X(z_1)}
e^{ik_2\cdot X(z_2)}
\cdots e^{ik_n\cdot X(z_n)}\rangle_{{\cal W}_\alpha,\, m}
\nn \\
&&=(2\pi)^D\delta^D(k_1 + k_2 + \cdots + k_n)
\left(\frac{\pi}{1+\alpha}\right)^{\sum_{i=1}^n k_i^2}
\prod_{1\le i<j}^n
\Big|\sin\left( \frac{\pi(z_i-z_j)}{1+\alpha}\right) \Big
|^{2k_i\cdot k_j},
\nn\\
\eear
where $\langle\cdot\rangle_{{\cal W}_\alpha,\, m}$ denotes the
matter correlator on the wedge state surface ${\cal W}_\alpha$.

\subsection{Calculation of $\beta_n^A$}

Using the formulas (\ref{BPZcor}), (\ref{COR3}), (\ref{Bcccc2}),
and (\ref{npc2}) defined in this Appendix, we can calculate
the tachyon coefficient $\beta_n$ in Eq.~(\ref{betn2}).
Firstly we calculate $\beta_n^A$ which comes from the contribution of
$Q_B$,
\bear\label{betnA}
\beta_n^A=\langle\chi_n,\, A_n\rangle,
\eear
where
\bear\nn
&&\chi_n= e^{nX^0(0)}c_0c_1\ket{0},
\nn \\
&&A_n =\left[ \partial\left(e^{-X^0(0)}c(0)\right)\ket{0}
+e^{-X^0(0)}c_1K_1^L\ket{0}\right]*J^{n-2}*e^{-X^0(0)}B_1^L c_1\ket{0}.
\eear
From the formula (\ref{BPZcor}) we obtain
\bear\label{betnA2}
\beta_n^A&=&\left(\frac{2}{\pi}\right)^{n^2+n-1}\Bigg[
\frac{\partial}{\partial x}\Bigg\{\langle\partial c(0) c(0) c(x)
{\cal B} c(n)\rangle_{{\cal W}_n,\, g}
\nn \\
&&\times\Big\langle e^{nX^0(0)} e^{-X^0(x)} e^{-X^0(2)} e^{-X^0(3)}
\cdots e^{-X^0(n)}\Big\rangle_{{\cal W}_n,\,m}\Bigg\}_{x=1}
\nn \\
&&+\frac{\partial}{\partial m}\Bigg\{\langle
\partial c(0) c(0) c(1) {\cal B} c(m+n-2)\rangle_{{\cal W}_{m+n-2},\, g}
\nn \\
&&\times \Big\langle e^{nX^0(0)} e^{-X^0(1)} e^{-X^0(m)} e^{-X^0(m+1)}
\cdots e^{-X^0(m+n-2)}\Big\rangle_{{\cal W}_{m+n-2},\,m}\Bigg\}_{m=2}
\Bigg],
\eear
where we used the Eq.~(\ref{BBB}) and the facts,
\bear\nn
f\circ c_0\ket{0} = \partial c(0)\ket{0},\quad
f\circ c_1\ket{0}= \left(\frac{\pi}{2}\right) c(0)\ket{0},\quad
K_1^L\ket{0} = \left(\frac{2}{\pi}\right)
\frac{\partial}{\partial m}\ket{m}|_{m=2}.
\eear
Using the translation symmetry on SIC, we can rewrite
the ghost correlators in Eq.~(\ref{betnA2}) as
\bear\label{Bcccc3}
&&\langle\partial c(0) c(0) c(x)
{\cal B} c(n)\rangle_{{\cal W}_n,\, g} =
\langle {\cal B} c(-1)\partial c(0) c(0) c(x)
\rangle_{{\cal W}_n,\, g},
\nn \\
&&\langle \partial c(0) c(0) c(1)
{\cal B} c(m+n-2)\rangle_{{\cal W}_{m+n-2},\, g}
=\langle {\cal B} c(-1)\partial c(0) c(0) c(1)
\rangle_{{\cal W}_{m+n-2},\, g}.
\eear
Then applying formulas (\ref{Bcccc2}), (\ref{npc2}), and
(\ref{Bcccc3}) to the Eq.~(\ref{betnA2}), we obtain
the explicit expression of $\beta_n^A$ in Eq.~(\ref{betnABC}).

\subsection{Calculation of $\beta_n^B$}

In the calculation of $\beta_n^B$ we use the tachyon vacuum solution
$\Psi$ given in Eq.~(\ref{SS}). Then $B_n$ in Eq.~(\ref{ABC}) is
rewritten as
\bear\label{Bn}
B_n&=&\Psi * \phi^n = \Psi * J^{n-1}*e^{-X^0(0)}B_1^Lc_1\ket{0}
\nn \\
&=& \sum_{m=0}^\infty\frac{\partial}{\partial m}\left[
\frac{2}{\pi} c_1\ket{0}*\ket{m+1}*J^{n-1}*
e^{-X^0(0)}B_1^Lc_1\ket{0}\right]
\nn \\
&&-\lim_{N\to\infty}\left[\frac{2}{\pi}
c_1\ket{0}*\ket{N+1}*J^{n-1}* e^{-X^0(0)}B_1^Lc_1\ket{0}\right].
\eear The tachyon vacuum solution is composed of the ordinary piece
$\sum_{n=0}^\infty\psi_n^{'}$ and the phantom piece $-\psi_\infty$.
However, the contribution of the phantom piece which corresponds to
the last term in Eq.~(\ref{Bn}) to $\beta_n^B$ vanishes since
\bear\nn \langle\chi_n,\,\psi_N*\phi^n\rangle\sim {\cal O}
\left(\frac{1}{N^3}\right) \eear for large $N$. By neglecting the
phantom piece in the calculation of $\beta_n^B$, we obtain
\bear\label{betnB} \beta_n^B&=& \left(\frac{2}{\pi}\right)^{n^2+n-1}
\sum_{m=0}^\infty\frac{\partial}{\partial m} \Big[\langle\partial
c(0) c(0) c(1) {\cal B} c(m+n+1) \rangle_{{\cal W}_{m+n+1},\,g}
\nn \\
&&\times \Big\langle e^{nX^0(0)} e^{-X^0(m+2)} e^{-X^0(m+3)}
\cdots e^{-X^0(m+n+1)}\Big\rangle_{{\cal W}_{m+n+1},\, m}\Big].
\eear
Similarly to the case of $\beta_n^A$, by using the translation symmetry
on SIC and the ghost and matter correlators (\ref{Bcccc2}) and
(\ref{npc2}) we can obtain the expression $\beta_n^B$ in
Eq.~(\ref{betnABC}).

\subsection{Calculation of $\beta_n^C$}

Similarly to the case of $\beta_n^B$ in the previous subsection,
by neglecting the phantom piece in the calculation of $\beta_n^C$
we obtain
\bear\label{betnC}
\beta_n^C &=& -\left(\frac{2}{\pi}\right)^{n^2+n-1}
\sum_{m=0}^\infty\frac{\partial}{\partial m}\Big[
\langle\partial c(0) c(0) {\cal B} c(1) c(2) {\cal B} c(m+n+1)
\rangle_{{\cal W}_{m+n+1},\, g}
\nn \\
&&\times \Big\langle e^{nX^0(0)} e^{-X^0(1)} e^{-X^0(m+3)}
e^{-X^0(m+4)}\cdots e^{-X^0(m+n+1)}\Big\rangle_{{\cal W}_{m+n+1},\, m}
\Big].
\eear
In the calculation of the ghost correlator in Eq.~(\ref{betnC}),
we use the relations $\{{\cal B},\, c(z)\}=1$, ${\cal B}^2=0$, and
the translation symmetry on SIC and obtain
\bear\label{ccBccBc}
&&\langle\partial c(0) c(0) {\cal B} c(1) c(2) {\cal B} c(m+n+1)
\rangle_{{\cal W}_{m+n+1},\, g}
\nn \\
&&= \langle {\cal B} c(-1)
\partial c(0) c(0) c(2)\rangle_{{\cal W}_{m+n+1},\, g}
-\langle {\cal B} c(-1)\partial c(0) c(0) c(1)
\rangle_{{\cal W}_{m+n+1},\, g}.
\eear
Using the relations (\ref{Bcccc2}), (\ref{npc2}), and
(\ref{ccBccBc}), we obtain the explicit expression of $\beta_n^C$
in Eq.~(\ref{betnABC}).

\end{document}